\documentclass[11pt]{article}
\usepackage[lmargin = 0.9in,
            rmargin = 0.9in,
            margin = 0.5in,
            margin = 1.0in]{geometry}
\usepackage{float}
\usepackage{endfloat}
\usepackage{threeparttable}
\usepackage{amssymb, amsmath}
\usepackage{graphics}
\usepackage{authblk}
\usepackage{graphicx}
\usepackage{enumitem}
\usepackage{hyperref}
\usepackage{url}
\usepackage[titletoc,toc]{appendix}
\usepackage{threeparttable}
\usepackage{longtable}
\RequirePackage{filecontents}
\usepackage{cleveref}
\usepackage{endfloat}
\usepackage{color,soul}
\usepackage{subfig}
\usepackage{natbib}

\newenvironment{acks}[1]%
{\subsection*{\normalsize\bfseries Acknowledgements}\noindent #1}

\begin{document}
\linespread{1.0}
\title{\textbf{Robust Design and Analysis of Clinical Trials With Non-proportional Hazards: A Straw Man Guidance from a Cross-pharma Working Group}}

\author[1]{Satrajit Roychoudhury}
\author[2]{Keaven M Anderson}
\author[3]{Jiabu Ye}
\author[4]{Pralay Mukhopadhyay}
\affil[1]{Pfizer Inc., New York, NY}
\affil[2]{Merck and Co, Inc, North Wales, PA}
\affil[3]{Astrazeneca Pharmaceuticals, Gaithersburg, MD}
\affil[4]{Otsuka America Pharmaceuticals, Inc., Rockville, MD}
\date{}
\maketitle

\newpage
\linespread{1.6}

\begin{abstract}
Loss of power and clear description of treatment differences are key issues in designing and analyzing a clinical trial where non-proportional hazard is a possibility. A log-rank test may be  inefficient and interpretation of the hazard ratio estimated using Cox regression is potentially problematic. In this case, the current ICH E9 (R1) addendum would suggest designing a trial with a clinically relevant estimand, e.g., expected life gain. This approach considers appropriate analysis methods for supporting the chosen estimand. However, such an approach is case specific and may suffer from lack of power for important choices of the underlying alternate hypothesis distribution. On the other hand, there may be a desire to have robust power under different deviations from proportional hazards. We would contend that no single number adequately describes treatment effect under non-proportional hazards scenarios. The cross-pharma working group has proposed a combination test to provide robust power under a variety of alternative hypotheses. These can be specified for primary analysis at the design stage and methods appropriately accounting for combination test correlations are efficient for a variety of scenarios. We have provided design and analysis considerations based on a combination test under different non-proportional hazard types and present a straw man proposal for practitioners. The proposals are illustrated with real life example and simulation.
\end{abstract}
\bigskip
\bigskip
\bigskip
\bigskip

\noindent {\it Key Words:}
Non-proportional hazards, Log-rank test, Weighted log-rank test, Combination test, Clinical trial design

\newpage


\section{Introduction}

A time to event endpoint is the primary outcome of interest in many clinical trials. For such trials, each subject is either experience an event of interest (e.g., disease progression or death) or censored. Commonly used statistical methods for comparing two survival curves in a randomized trial are the Kaplan-Meier survival plot (\cite{KaplanMeier}), log-rank test (\cite{petologrank}), and Cox regression (\cite{Cox1972}). The performance of the log-rank test and Cox regression heavily depend on the proportional hazards (PH) assumption. In reality, the PH assumption is often not met. For example, recent immuno-oncology therapies pose unique challenges to the trial design where a delayed separation of the Kaplan-Meier curves have often been observed, potentially resulting in a violation of proportional hazards (PH) assumption (~\cite{Chen2013}, \cite{Kaufman2015}, \cite{Borghaei2015}, \cite{Herbst2016}).   
     
Since treatment differences under non-proportional hazard (NPH) constitute a broad class of alternative hypotheses, finding one test and estimate of treatment benefit that are consistently meaningful and provide good statistical power in multiple situations is challenging. Well known NPH types observed in clinical trials are {\em delayed effect}, {\em crossing survival}, and {\em diminishing effect} over time. While there may be speculation about the nature of treatment effect at the time of study design, we have found that it can often be wrong. This further complicates the trial design and treatment effect quantification. Therefore, a suitable design and analysis method for time to event data with potential NPH must be able to describe multiple alternatives in a meaningful way as well as provide competitive power to optimal tests across many scenarios. In addition, a single treatment effect summary is not adequate to capture the time dependent nature of the benefit. We need measures beyond a simple hazard ratio or restricted mean survival time (RMST) to quantify and communicate the treatment effect.

The goal of this paper is to guide practitioners about strengths and limitations of the available methods of design and analysis of clinical trial with potential NPH. We evaluate them as a candidate for primary analysis in the confirmatory studies.  These are meant as straw man proposals for initiating a general discussion with different stakeholders. While considerable thought and background investigation has gone into this proposal by a cross-pharma working group, it is not endorsed by regulatory authorities. Project specific customization is necessary to fulfill the needs in a particular situation.     

The rest of the paper is structured as follows. We start with a brief overview of the statistical methods available in the literature and their merit as the primary analysis in confirmatory trials. This is followed by some recommendations for analysis and design approaches when NPH is plausible. Both the sections are complemented with illustrative examples. Discussion and concluding remarks are in the last section.

\subsection{Overview of Available Methods}\label{sec:overview}
There is a vast literature available on the analysis of time to event data in presence of NPH. In this paper, we mainly focus on the statistical methods related to drug development. Till date, the log-rank (LR) test is the regulatory accepted standard test for comparing two survival curves.  Being  nonparametric in nature, the LR test is statistically valid when the PH assumption is not met, but it has poor power in certain situations (~\cite{Chen2013}). Moreover, the hazard ratio (HR) generated by the Cox proportional hazards model has limited interpretation as a treatment effect summary when data fails to support the PH assumption. Possible alternative is a general class of weighted log-rank test (WLR) introduced by Fleming-Harrington (FH) ($G_{\rho,\gamma}$) (~\cite{FlemingHarrington1982}) and weighted hazard ratio (WHR) (~\cite{FlemingHarrington1982, Schemper1992}). Other available options are piecewise LR test (~\cite{Xu2017, Xu2018}), modestly weighted LR test (mWLRT) (~\cite{MagirrBurman2018}) and the piecewise HR. For rank based methods other than LR test, model parameters (e.g., weight function, intervals) play an important role in the successful implementation of all these methods. Therefore, achieving robust power under wide number of NPH alternative is difficult.

Recently, Kaplan-Meier (KM) based methods such as difference in milestone survival rates at fixed time points (~\cite{Klein2007}) and restricted mean survival time (RMST) (~\cite{RoystonParmar2011, Uno2014, Tian2014}) have become popular to the non-statisticians for its intuitive and simplistic interpretation under NPH. Weighted Kaplan-Meier test (WKM) (~\cite{PepeFleming1989}) and  restricted mean time lost (RMTL) are the other methods in the same class. Performance of KM based methods depend on the length of study period ($\tau$) and censoring pattern. Simulation studies have shown that the power gain for the WKM and RMST are minimal in comparison to the LR test when there is a delayed effect (~\cite{NPHsim2018}). Moreover, the time point(s) for milestone survival analysis needs to be chosen carefully for proper clinical interpretation. 

Cox regression has been widely used in clinical trial for treatment effect estimation. A natural extension of Cox regression model for NPH setting is including a time varying coefficient for treatment (CoxTD) (~\cite{Putter2005}). However, one major challenge of this approach is the specification of appropriate ``time function" before trial begins. ~\cite{Putter2005} also suggested ``log-time" function as a ``reasonable" choice to diminish the influence of very early events. An extensive simulation study in ~\cite{Callegaro2017} showed that the CoxTD model does not perform well in terms of power when the underlying survival pattern is delayed treatment effect. Moreover,  the HR as a continuous function of time for the primary treatment effect summary is not appealing to clinicians, regulators and patients.

Combination tests provide robust testing of hypothesis approach when the type of NPH is unknown. In a combination test, multiple test statistic are considered at the same time to handle possibilities of different types of NPH when trial data emerges. A combination test can combine multiple rank-based tests (~\cite{Lee2007}) or rank-based tests with KM based tests (~\cite{Chi2000}). Notable works in this area include ~\cite{Breslow1989}, ~\cite{Lee1996}, ~\cite{Chi2000}, ~\cite{Lee2007}, ~\cite{Logan2008}, ~\cite{YangPrentice2010}, ~\cite{Karrison2016}, ~\cite{Callegaro2017}, and ~\cite{NPHsim2018}. Moreover, combination tests do not require pre-specification of a unique test  while designing the trial. Due to the flexibility of the test statistic, the combination test often provides robust power under wide class of alternative hypotheses. We have discussed utilities of one particular combination test further in Section 2.1.1 and 2.1.2.    

Other important work in this area are accelerated failure time (AFT) model (~\cite{James1979, Prentice1978, Chapman2013}) and net benefit  (~\cite{Buyse2010, Paron2016}). Further details of   different approaches are provided in the supplementary appendix.
  
\section{Analysis Approaches of Clinical Trials with Non-proportional Hazards}

A key challenge is to specify the primary analysis of a clinical trial when there are considerable uncertainties regarding  the actual NPH type. Recent trials show that a delayed effect is common in immuno-oncology (~\cite{Chen2013}). Other NPH types such as crossing survival and diminishing effects have been also observed in different therapeutic areas. At the time of trial design, PH, a delayed effect with unknown delay or even crossing hazards with an uncertain degree and timing of crossing can all be plausible. A good primary analysis method for trials with potential NPH needs to take into account all the uncertainties mentioned above to provide robust statistical inference. In this section, we propose a few candidates for testing the primary endpoint in a confirmatory trial along with relevant treatment effect quantifiers. We provide three real life examples for illustration.

\subsection{Choice of Primary Analysis Method for a Confirmatory Trial} 

Based on the ICH E9 guideline (~\cite{ICHE9}), the primary analysis of a trial needs to be planned prior to enrolling patients. This increases the degree of confidence in the final results and conclusions of the trial. A primary analysis involves both testing and estimation of treatment effect that goes to the label of a drug if approved. With potential NPH, it is difficult to specify a statistical method that can provide consistently high power across PH and different NPH scenarios.

At first, we perform a qualitative evaluation of the methods described in Section ~\ref{sec:overview} as  possible candidates for the hypothesis testing in the primary analysis of a confirmatory trial. The purpose of this comparison is to help statisticians understand the merits and shortcomings of each  method as a candidate for the primary analysis. We consider the following four metrics as the basis for this comparison. Table ~\ref{tab:NPH:Methods} summarizes the advantages and disadvantages of each approach.

\begin{enumerate}
\item Type I error: Controlling type I error at a specific level of significance (e.g., 2.5\%) under the null hypothesis $H_{0}$: $S_{C}(t)$ = $S_{T}(t)$ for all $t$. Here $S_{C}(t)$ and $S_{T}(t)$ are the underlying survival functions for control and treatment group respectively.    
\item Robust power: Showing resilience in terms of statistical power when the PH assumption is violated. Often a statistical test suffers a power loss when the nature of the underlying treatment effect is not anticipated. 
\item Treatment effect Interpretation: Interpretable treatment effect summary under various types of PH and NPH   
\item Non-statistical Communication: Easy to understand by non-statisticians
\end{enumerate}

\begin{table}[ht]
\caption{Review of Different Approaches for Analysis of Time to Event Data Under NPH} \label{tab:NPH:Methods}
\centering
\smallskip 
\begin{tabular}{lcccc}
\hline
Method                               & Type I   & Robust &  Treatment Effect &  Non-statistical\\
                                     & Error    & Power  &  Interpretation   & Communication  \\
\hline
\textbf{Rank-based Test}                                 &      &     &      &      \\
                                                         &      &     &      &       \\
\hspace{2mm} Log-rank Test/Cox Model                     & Yes  & No  & No   &  Yes \\
\hspace{2mm} Fleming Harrington weighted                 & Yes   & No  & No   &   No  \\
\hspace{2mm} Log-rank Test/Weighted Hazard Ratio         &      &     &      &       \\
\hspace{2mm} Piecewise Log-rank Test/                    & Yes  & No  & Yes  &  Yes \\
\hspace{2mm} Modestly weighted Log-rank Test/            &      &     &      &      \\
\hspace{2mm} Piecewise Hazard Ratio                      &      &     &      &      \\
\hline
\textbf{Kaplan-Meier based Test}                         &      &     &      &      \\
\hspace{2mm} Milestone Survival                          &  Yes & No  &  Yes &  Yes  \\
\hspace{2mm} Kaplan-Meier Median                         &   -  & -   &  No  &  Yes \\
\hspace{2mm} Weighted KM test                            &  Yes & No  &  No  &   No \\
\hspace{2mm} RMST                                        &  Yes & No  &  Yes &  Yes \\
\hline
Cox model with time varying                              &  Yes & No  &  Yes  & No  \\
treatment effect (CoxTD)                                 &      &     &       &     \\
\hline
 Combination test                                        &  Yes & Yes &  Yes &   No \\
\hline                                                   &      &     &      &       \\
Net Benefit                                              &  Yes &  No &  Yes &  Yes \\
\hline
\end{tabular}
\end{table}

Based on the assessment above WKM, milestone survival, RMST, CoxTD, and combination tests are potential candidates for hypothesis testing method in the primary analysis of a confirmatory trial when NPH is expected. But, WKM, CoxTD, milestone survival and RMST fail to show robust power under a wide class of alternatives (~\cite{NPHsim2018}, ~\cite{Callegaro2017}). In the next subsection we will introduce a new combination test for confirmatory trials which is an improvement over the available tests and provides robust power under various NPH scenarios.  If NPH is not expected, we recommend the use of traditional LR test and HR for the primary analysis.  

\subsubsection{Robust MaxCombo Test}
We propose a new combination test as an alternative choice for primary analysis of a confirmatory trial when NPH is a possibility.  The test is based on multiple Fleming-Harrington WLR test statistics and chooses the best one adaptively depending on the underlying data. The main objective of this test is to provide robust power for primary analysis under different scenarios.  We refer to it as the {\em MaxCombo} test. This idea is motivated from the work by ~\cite{YangPrentice2010} and ~\cite{Lee2007}. 

The MaxCombo test considers the maximum of four correlated Fleming-Harrington WLR test statistics; $G^{0,0}$ (LR test), $G^{0,1}$, $G^{1,1}$, and $G^{1,0}$ (Prentice-Wilcoxon) together provides a robust test under different scenarios including PH, delayed effect, crossing survival, early-separation, and mixture of more than one NPH type scenarios as an alternative. Other weight function for WLR test can also be considered depending on the expected outcome. We propose alternate weights as another option in the next sub-section. By construction MaxCombo is less dependent on the underlying shape of the survival curve than a single WLR and shows good power across many alternatives with respect to the optimal design for each alternative. 

The type I error and power calculation require the joint distribution of four WLR statistics. Under the null hypothesis, joint distribution of $G^{0,0}$, $G^{0,1}$, $G^{1,1}$, and $G^{1,0}$ asymptotically follows a multivariate normal distribution (~\cite{Karrison2016}). The p-value of MaxCombo test is calculated by using a 4-dimensional multivariate normal distribution. This calculation can be done using efficient integration routine in standard statistical software like R and SAS (~\cite{Ganz1992}). Further details of the variance-covariance calculation are provided in Appendix A. MaxCombo is flexible enough to incorporate other weighting schemes as well. However,  the variance-covariance structure may not be in closed form for other weight choices and requires intensive computation. 

An extensive simulation study under the null hypothesis and different treatment effect scenarios was performed by the cross-pharma working group (~\cite{NPHsim2018}) to understand the statistical operating characteristics (type I error and power) of MaxCombo test. The simulation study considered varied enrollment patterns, number of events, and total study duration. It showed that the type I error is well protected below 2.5\% under the null hypothesis (no treatment effect). The simulation study also demonstrated robust power of the MaxCombo test under different alternatives. It showed a clear benefit over LR, WKM and RMST in terms of power in presence of the delayed effect and crossing survival. The proposed MaxCombo also showed benefit over the ~\cite{Lee2007} combination test when the underlying hazard is delayed effect with converging tails (\cite{NPHsim2018}). The Lee's method (~\cite{Lee2007}) is computationally intensive than MaxCombo due to the use of simulation. Moreover, the power loss of the MaxCombo test was minimal (3-4\%) as compared to the LR test when the underlying treatment effect is PH. In summary, the MaxCombo test fulfills the necessary regulatory standards and is suitable for a confirmatory trial with potential NPH.

\subsubsection{Additional Investigation of MaxCombo Test}

We have further investigated the properties of the MaxCombo test in extreme scenarios such as  {\em strong null} (~\cite{MagirrBurman2018, FreidlinKorn2019}) and {\em severe late crossing}. A {\em strong null} scenario refers to a situation when the survival distribution of treatment group is uniformly inferior to the control group (i.e., $S_{C}(t) \; \geq \; S_{T}(t)$ for all $t$). Two different {\em strong null} scenarios are considered here. The First scenario is from the ~\cite{MagirrBurman2018} paper and referred as {\em strong null 1}. We have also explored another extreme scenario ( referred as {\em strong null 2}) which is adopted from the ~\cite{FreidlinKorn2019}. For the {\em severe late crossing} scenario, the treatment group shows a late and marginal survival benefit  over the control group which makes the overall treatment effect clinically questionable. Each scenario is evaluated with 20,000 iterations. This half-width of the simulation 95\% confidence interval for 2.5\% type-I error and 80\% power are 0.002 and 0.0055 with 20,000 iterations. This level of precision is considered sufficient to understand the simulation results. 

Following ~\cite{MagirrBurman2018}, we have assumed a two-arm randomized control trial with 100 patients per arm, recruited at a uniform rate over 12 months. For the control arm, survival data is generated using an exponential distribution with a median of 15 months. The final cut-off date for each simulation is the calendar time of 36 months after the start of the study. All patients alive at that point are censored at the cut-off date. The following scenarios are considered for the treatment arm using two piece exponential distribution:
\begin{enumerate}
	\item[a)] Strong null 1: For the first 6 months, the hazard rate is $\log 2 /9$ $=$ 0.077 (i.e, hazard ratio 1.67 (treatment vs control)).  After 6 months, the hazard rate is approximately 0.04 (i.e, hazard ratio 0.87 (treatment vs control)). This ensures that the survival probability for patients in control arm is always better than in treatment arm. The curves meet at the 36 month.
	months.
	\item[b)] Severe late crossing: For the first 6 months, the survival rate is $\log 2 /9$ $=$ 0.077 (i.e, hazard ratio 1.67 (treatment vs control)).  After 6 months the hazard rate is approximately 0.036 (i.e, hazard ratio 0.79 (treatment vs control)). This ensures that the survival probabilities for patients in control arm is better than in treatment arm till month 27. The survival probability of the treatment arm is marginally better than control arm afterwards. However, the treatment effect in this scenario is not clinically relevant.  
\end{enumerate}

For the strong null 2, we have assumed two-arm randomized control trial with 1000 patients per arm, recruited instantaneously. For the control arm, survival data is generated using an exponential distribution with a median of 2.7 years (i.e., constant hazard rate 0.25). The final cut-off date for each simulation is the calendar time of 5 years after the start of the study. All patients alive at that point are censored at the cut-off date. For the treatment arm, survival data is generated using a two piece exponential distribution. For the first 1.2 months, the hazard rate is $\log 2 /0.1732868$ $=$ 4 (i.e, hazard ratio 16 (treatment vs control)).  After 1.2 months, the hazard rate is approximately 0.19 (i.e, hazard ratio 0.76 (treatment vs control)). This set up is consistent with the one proposed by  ~\cite{FreidlinKorn2019}. 

\begin{figure}
	\centering
	\subfloat[Strong null 1]{\includegraphics[width=0.4\textwidth]{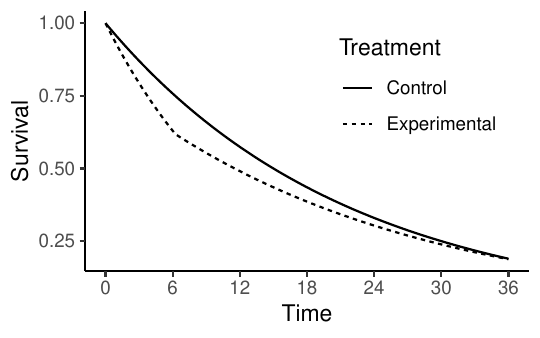}}
    \subfloat[Strong null 2]{\includegraphics[width=0.4\textwidth]{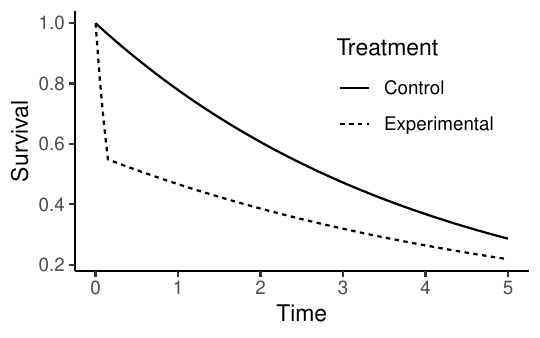}}
	\hfill
	\subfloat[Severe late Crossing]{\includegraphics[width=0.4\textwidth]{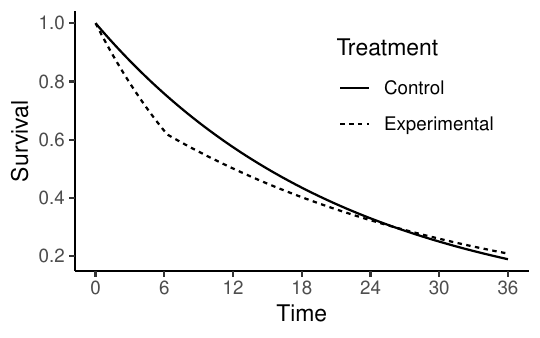}}
	\caption{Simulation scenarios for evaluation of MaxCombo under strong null and severe late crossing}
	\label{fig:MaxCombosim}
\end{figure}

The underlying survival distributions of treatment and control arms under two strong null and severe late crossing scenarios are shown in Figure ~\ref{fig:MaxCombosim}. We presume that the trials run with either of these scenarios would most likely be stopped early by a data monitoring committee (DMC) due to the safety concerns if these scenarios are observed in real life (~\cite{FreidlinKorn2009}). {\em Strong null} situations are unlikely in the confirmatory trials. Therefore, these findings should not be mixed with type-I error assessment. 

Simulations show that the probability of rejecting null hypothesis is as low as 2.1\% for the MaxCombo test under the strong null 1 scenario due to the multiplicity adjustment. For the severe late crossing scenario, the probability of rejecting null hypothesis is also small (5.0\%). The results are similar if the recruitment period is 6 months (strong null 1: 2.3\% and severe late crossing: 5.8\%). This further confirms favorable operating characteristics of the MaxCombo test under these extreme scenarios. However for the strong null 2,  the probability of rejecting null hypothesis goes up to 48.9\% due to it's extreme nature.

If extreme scenarios like severe late crossing and strong null 2 are of major concern, one can consider an alternatives weights for WLR in MaxCombo test. For example, we have investigated a modified MaxCombo test with $G^{0,0}$, $G^{0,0.5}$, $G^{0.5,0.5}$, and $G^{0.5,0}$. The probabilities of rejecting null hypothesis with modified MaxCombo are reduced significantly to  2.6\% under the severe late crossing scenario and 1.8\% under the strong null 2. The power loss is also minimal as compared to the original MaxCombo test. Furthermore, we looked into the proportional hazard (PH) and delayed effect (DL) defined by  ~\cite{MagirrBurman2018} to understand the impact on power. The power of the modified MaxCombo test is comparable with the original MaxCombo test. For PH and DL scenarios; the powers for modified MaxCombo test are 76.1\% and 78.2\%  respectively. The powers for original MaxCombo test are 74.4\% and 79.9\%  respectively. The results are still superior to the LR and modestly weighted LR test (~\cite{MagirrBurman2018}). Therefore, the modified MaxCombo test with more moderate down-weighting is a good alternative if practitioners want to have strict control for extreme scenarios like severe late crossing and strong null 2. Note that, both statistically significant and clinically relevant results are required for regulatory approval, regardless of the testing strategy used.

\subsection{Specification of the Primary Analysis}
The primary analysis involves both hypothesis testing and estimation of treatment effect. Under NPH, a single summary statistic measure (e.g. HR or RMST) fails to capture the time dependent treatment effect and is heavily dependent on the follow-up duration. Therefore, multiple treatment effect summaries are critical to summarize and understand the overall risk-benefit profile. Also, it is important that sufficient follow-up is available to characterize both short- and long-term effects. We propose a three step approach for primary analysis using the MaxCombo test when there is a chance of observing NPH. The main goals of the proposed approach are the use of a robust test and provide appropriate treatment effect summaries. In spite of the early specification, the proposed approach reports the {\em best} treatment effect summaries in adaptive manner based on the data. Similar approaches based on the LR test and CoxTD were discussed by ~\cite{RoystonParmar2011} and ~\cite{Campbell2014}. However, these approaches are less efficient due to a test with low power for important scenarios. It is important to note that this approach is a shift from the traditional paradigm where a dual estimator corresponding to a primary testing procedure is always reported as the primary treatment effect quantifier. For example, the HR from the Cox regression model is always presented as the dual treatment effect quantifier for the LR test. 

\begin{itemize}
\item \underline{Step 1 (Test of null hypothesis):} This step is the hypothesis testing part of  primary analysis. Treatment effect should be tested with the MaxCombo test (or modified MaxCombo test) and conclusion regarding the null hypothesis is drawn accordingly. If the MaxCombo test is not significant, one concludes that the benefit of experimental treatment has not been demonstrated.
 
\item \underline{Step 2 (Assessment of PH):} Regardless of the step 1, the PH assumption of the underlying treatment effect needs to be assessed based using different tools like Grambsch-Therneau test or G-T test (~\cite{GT1994}) based on the scaled Schoenfeld residuals from a Cox model and other visual diagnostics (KM plot, hazard plot, log-log survival plot).
\item \underline{Step 3 (Treatment effect summary):} The treatment effect quantifiers will depend on the outcome of Step 2
\begin{itemize}
\item If PH assumption is reasonable: report the HR from Cox regression and corresponding 95\% confidence interval (CI) as the primary treatment effect measure. The milestone survival rates are also useful treatment effect quantifiers in this context.
\item If PH assumption is not reasonable: In this situation, one summary measure fails to explain the overall treatment effect. We recommend presenting more than one treatment effect quantifiers, such as:
\begin{itemize}
\item Ordinary HR estimate (treatment vs control) and 95\% CI from the Cox regression
\item Difference in milestone survival rate (treatment vs control) and 95\% CI at clinically relevant time point $t^{*}$ 
\item Difference in RMST (treatment vs control) at $t^{*}$ and 95\% CI: gain in life expectancy at the minimum of the maximum observed survival in treatment and control group
\end{itemize}
\end{itemize}
\end{itemize}

We also recommend to report the Kaplan-Meier plot, the milestone survival rates at additional time points, the piecewise HR and the net benefit as supportive measures. These measures are useful to understand the time dependent treatment profile and communicate to non-statisticians easily. Alternatively, one can use CoxTD and provide the time dependent HR along with 95 \% CI (\cite{Putter2005}).   

For completeness, one can also look into the WHR estimate (treatment vs control) and 95\% CI using the best weight chosen by MaxCombo and using weighted Cox regression. It is the dual measure of MaxCombo test. Alternatively, one can report average hazard ratio proposed by ~\cite{Schemper2009} and ~\cite{Xu2000}. However, the WHR has limited interpretation to non-statisticians. Similar to the p-value, the 95\% CI corresponding to the estimated WHR using the best weight from MaxCombo test needs to be adjusted due to the positive correlation between four WLR test statistics. A simultaneous confidence interval procedure by using asymptotic multivariate normal joint distribution of WHR is provided in Appendix B. 

Given the complexity of the NPH (e.g., delayed effect, crossing survival), the p-value from step 1 or a single summary statistic fails to capture the treatment benefit. Especially, scenarios with crossing survival require a careful evaluation and consideration of other factors such as the timing of crossing, treatment effect after crossing, and overall risk-benefit profile.  Due to the late emerging nature of the treatment effect, it is critical to follow patients for sufficient time to get reliable estimate of long-term benefit. In the unlikely extreme scenarios (e,g, severe late crossing in subsection 2.1.2), totality of the data will not support the approval of a new medication to the market despite a positive p-value. Our proposed step-wise approach chooses the appropriate treatment summary measures based on the situation and provides a totality of evidence for making optimum decision for an experimental drug.

\subsection{Examples}

We have provided three published clinical trial examples to demonstrate the utility of the MaxCombo test and the proposed primary analysis strategy in confirmatory trials. These examples are based on enrollment, follow-up assumptions, and high early event rate that are commonplace for trials in metastatic cancer. The examples include scenarios with crossing survival,  delayed effect, and proportional hazards. The purpose of this exercise is to demonstrate the utility of the proposed approach for primary analysis to the practitioners. We have no intention to comment or judge the clinical activity of the treatments involved or any related regulatory decisions.

The first example is the Phase III trial (IM211) of atezolizumab vs chemotherapy in patients with advanced or metastatic urothelial carcinoma (~\cite{Powles2018}). For this exercise we have considered the overall survival (OS) endpoint for the patients in the intention-to-treat population and  with $>$ 1\% tumour-infiltrating immune cells (PD-L1 expression: IC1/2/3). Using ~\cite{Guyot2012}, the data for analysis is reconstructed from the KM plot published in the supplementary materials (page 20) of ~\cite{Powles2018}.  The survival curves cross between 4 and 5 months and show survival benefit of patients in atezolizumab arm. However, a stratified Cox regression analysis in the publication shows a non-significant treatment effect (HR=0.87, 95\% (0.71, 1.05)). Moreover the median OS is 8.9 months for atezolizumab as compared to 8.3 months for chemo, a difference of only 0.6 months.  

A second example in recurrent or metastatic head-and-neck squamous cell carcinoma  is the phase III (KEYNOTE 040) trial of Pembrolizumab versus standard-of-care therapy (~\cite{Cohen2019}). The primary endpoint of this trial was OS in the intention-to-treat population. Based on the KM plot published in ~\cite{Cohen2019}, the survival benefit of Pembrolizumab  has emerged after 5 months. Therefore, we consider this as an example for {\em delayed treatment effect}. Though the survival benefit after 5 months looks promising, a traditional stratified LR have produced a marginally nominal significant p-value (one-sided) = 0.016. A stratified Cox regression analysis shows a HR= 0·80 ; 95\% CI (0·65–0·98). 

The final example satisfies the PH assumption. It is a phase III clinical trial (PA3) of Erlotinib Plus Gemcitabine vs Gemcitabine alone in patients with advanced pancreatic cancer (~\cite{Moore2007}). The primary endpoint of the trial was OS. Similar to the IM211 trial, the analysis data set is reconstructed from the  KM curve published in ~\cite{Moore2007}. A stratified analysis shows prolonged survival on the erlotinib plus gemcitabine arm with a HR of 0.82 (95\% CI: (0.69, 0.99) and nominal statistically significant LR test (2-sided p-value =0.038). However, the difference in median survival was just 0.3 months, 6.24 months in the gemcitabine plus erlotinib group compared to 5.91 months in the gemcitabine alone group. The estimated effect size poses some question related to clinical relevance of the combination in terms of survival benefit.

\begin{figure}
	\includegraphics[width=\linewidth ]{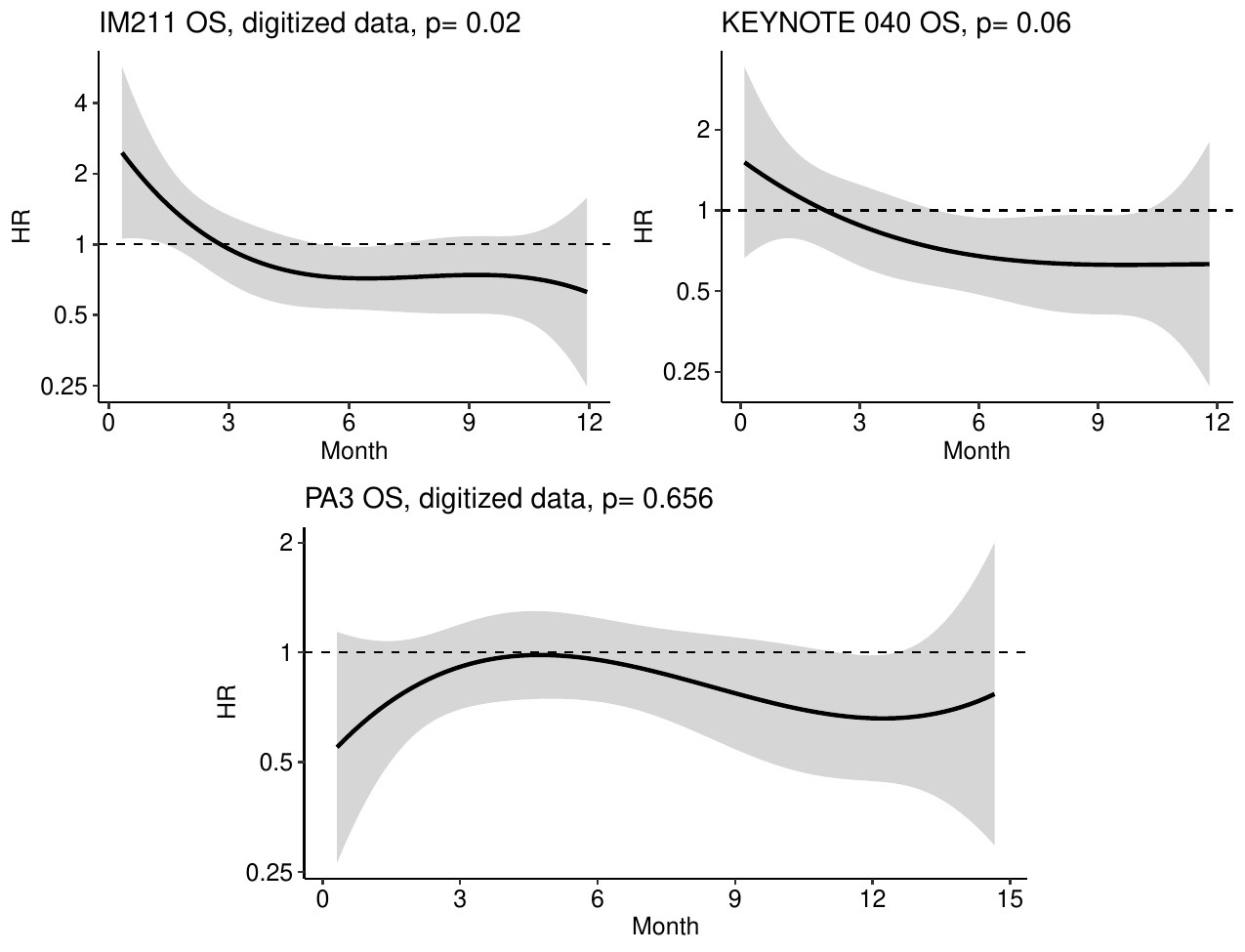}
	\caption{Schoenfeld Residual Plots for Three Examples:  a) IM211: Digitized (top left), b) KEYNOTE 040 (top right),  c) P   A3: Digitized (bottom)}
	\label{fig:EXSch}
\end{figure}

We start our analysis with the assessment of PH assumption for the three examples. The estimated hazard ratio over time is plotted along with the G-T test to ensure viability of the PH assumption (Figure ~\ref{fig:EXSch}) for three examples. A horizontal line represents a constant treatment effect. As expected, the PH hazard assumption is doubtful except for the PA3 trial. Both graphical diagnostics and G-T test fail to support the PH assumption for both IM211 and KEYNOTE 040 trials.  

Therefore, the traditional summary measures (e.g., Median, HR) fail to measure the treatment effect fully except for the PA3 trial. We have analyzed all the examples with MaxCombo, rank-based tests (LR and WLR), KM based tests (WKM, RMST, and RTML), difference in milestone rates at month 12 (clinically relevant time point), and \% net benefit for $>$ 6 months longer survival. For KM based methods, we have minimum of the largest observed time in each of the two groups as a choice of $\tau$. For nominal statistical significance, the p-value from each test is compared at one-sided 2.5\% level of significance. In addition, we have calculated the milestone survival rates at additional time points (at 3, 12, 18 and 24 months) and piecewise HR (intervals: 0- 3 months, 3- 6 months, 6-12 months and $>$ 12 months) to understand the treatment effect over time. Table ~\ref{tab:NPH:EXMethods} summarizes the analysis of OS for IM211, KEYNOTE 040, and PA3. As data set for two of the examples are re-constructed from published KM plots, we do not have the relevant stratification factors used for publication. Therefore, all the analyses in this section  are unstratified.

\begin{table}[ht]
	\caption{Analysis of Overall Survival for IM211 (Digitized), KEYNOTE 040, and PA3 (Digitized): Results of Different Methods} \label{tab:NPH:EXMethods}
	\centering
	\smallskip 
	\begin{threeparttable}
		\begin{tabular}{lccc}
			\hline
			Method                                 &   IM211: Digitized    &  KEYNOTE 040      & PA3: Digitized\\
			&     (Crossing)        &  (Delayed Effect) &  (PH)  \\
			\hline 
			Kaplan-Meier Median (months)           &                       &                   &        \\
			\hspace{2mm}  Treatment                &     8.9               &      8.4          &  6.24  \\
			\hspace{2mm}  Control                  &     8.3               &      6.9          &  5.91  \\
			\hline \\
			Log-rank Test                          &                       &                    &        \\
			\hspace{2mm}  p-value                  &     0.040             &       0.007        &  0.023  \\ 
			\hspace{2mm}  Cox HR                   &     0.847             &       0.778        &  0.834  \\ 
			\hspace{2mm}  95\% CI                  &   (0.70, 1.02)        &    (0.64, 0.95)    & (0.70, 0.99)\\ 
			\hline\\       
			Fleming Harrington WLR Test: p-value   &                       &                    &        \\
			\hspace{2mm} $G^{1,0}$                 &     0.216             &       0.068        &  0.047 \\
			\hspace{2mm} $G^{1,1}$                 &     0.004             &       0.001        &  0.064  \\
			\hspace{2mm} $G^{0,1}$                 &     0.002             &       0.001        &  0.031  \\
			\hline\\
			Max Combo                              &                       &                    &        \\
			\hspace{2mm}  Test selected            &     $G^{0,1}$         &     $G^{0,1}$      &  $G^{0,0}$ \\
			\hspace{2mm}  p-value                  &     0.005             &       0.001        &  0.048 \\
			\hspace{2mm}  Weighted HR              &     0.731             &       0.681        &  0.834  \\
			\hspace{2mm}  95\% CI                  &   (0.57, 0.93)        &   (0.52, 0.89)     & (0.68, 1.03) \\
			\hline\\
			RMST                                   &                       &                    &        \\
			\hspace{2mm}  Difference               &     1.090             &      1.900         &  0.860 \\
			\hspace{4mm}  95\% CI                  & (-0.22, 2.40)         &    (0.39, 3.41)    & (-0.07, 1.79)\\
			\hspace{4mm}  p-value                  &    0.051              &       0.007        &  0.034  \\
			RTML                                   &                       &                    &        \\
			\hspace{2mm}  Ratio                    &      0.920            &      0.891         &   0.942 \\
			\hspace{4mm}  95\% CI                  &   (0.83, 1.02)        &    (0.81, 0.98)    & (0.88, 1.01) \\
			\hspace{4mm}  p-value                  &      0.052            &       0.008        &   0.036    \\
			Weighted KM Test: p-values             &      0.129            &       0.024        &   0.034     \\
			\hline\\
			\% Net Benefit at Month 6              &                       &                    &        \\
			\hspace{2mm}  Difference               &       0.052           &      0.085         &  0.086  \\
			\hspace{4mm}  95\% CI                  &      (-0.04, 0.14)    &  (-0.014;0.19)     &(-0.02, 0.18) \\
			\hspace{4mm}  p-value                  &        0.286          &     0.106          &  0.083      \\
			\hline
		\end{tabular}
	\end{threeparttable}
\end{table}

\begin{figure}
	\includegraphics[width=\linewidth]{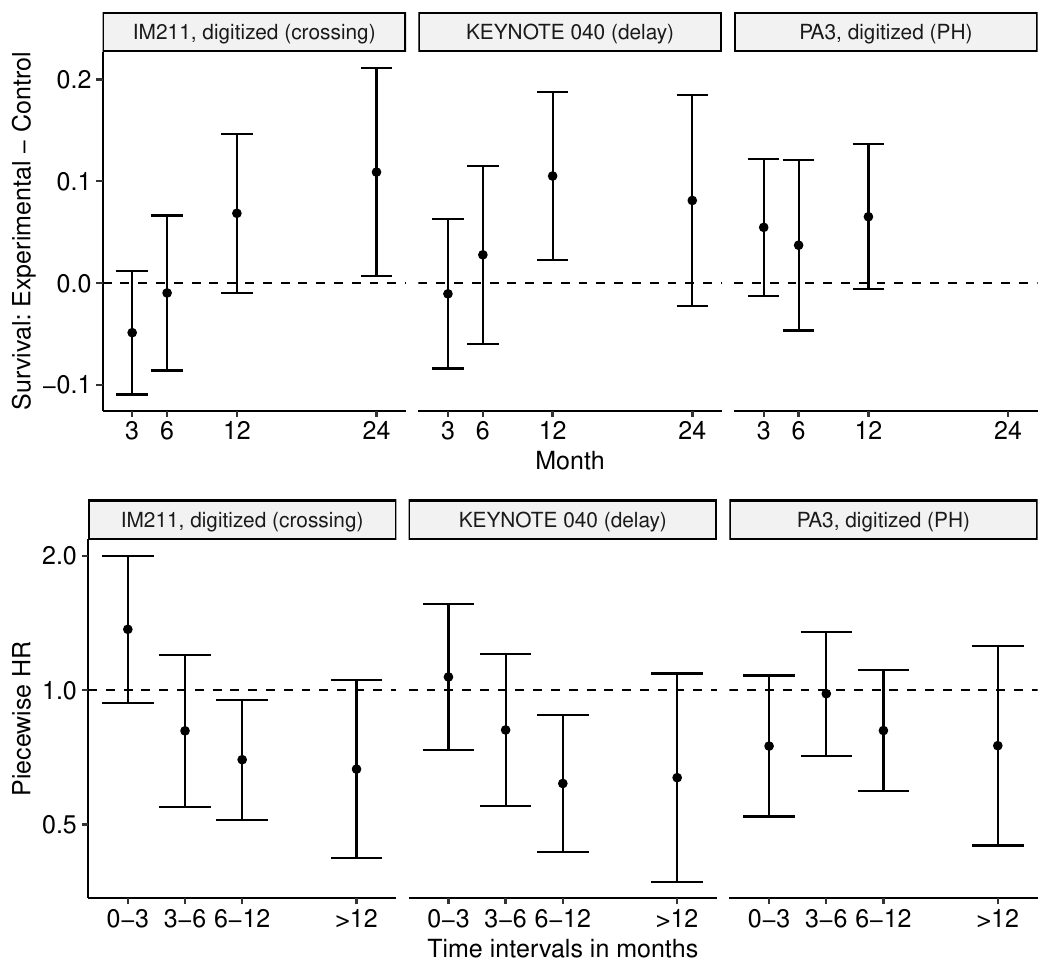}
	\caption{Milestone Survival (95\% CI) and Piecewise Hazard Ratio (95\% CI) at Clinically Relevant Time points for Three Trials}
	\label{fig:EX2}
\end{figure}

For the crossing survival scenario in IM211, Table ~\ref{tab:NPH:EXMethods} shows nominal significant result for the MaxCombo test (p-value= 0.005) only. All other tests fail to reject the null hypothesis in spite of potentially clinically relevant improvement of OS at the later stage. Therefore, one can miss a potential regulatory submission opportunity if the primary analysis is the LR test or other KM based tests. The MaxCombo chooses $G^{0,1}$ as the minimum p-value. As the diagnostic plot and result from the GT test (p value=0.02) questions the PH assumption, we recommend reporting the HR (estimate= 0.847, 95\% CI= (0.70, 1.02)), difference in milestone rates at 12 months (estimate= 0.021, 95\% CI = (-0.04, 0.18)) and difference in RMST (estimate= 1.09, 95\% CI=(-0.22, 2.40))) as treatment effect quantifiers in this case. HR and difference in RMST show a positive trend in favor of treatment. The overall evidence supports a clinically relevant treatment effect. Additional supportive analyses (e.g., milestone survival rate estimates at multiple timepoints, piecewise HR, WHR and \% net benefit) are useful to provide further confirmation of treatment benefit. Figure ~\ref{fig:EX2} shows the milestone survival estimates and piecewise HR's. Both capture the time-dependent treatment effect in an efficient manner. Difference in milestone survival rates at early and late timepoints show a survival benefit of atezolizumab over chemotherapy.

In KEYNOTE 040 trial, the treatment effect emerges late. Note that the published LR test p=0.016 was from a stratified analysis as compared to p-value=0.007 reported in Table ~\ref{tab:NPH:EXMethods} from a LR test without stratification. Similarly, stratified results for the MaxCombo test yields p-value =0.003 compared to a unstratified MaxCombo test with p-value =0.001 as reported in Table ~\ref{tab:NPH:EXMethods}. Thus, the stratified test was particularly important to clarify nominal statistical significance compared to the stratified LR test. Based on diagnostics plots (Figure ~\ref{fig:EXSch}) and GT test, there are some evidence of NPH. The HR over time plot challenges the constant HR assumption. Therefore, we have applied two strategies to understand the robustness of treatment effect. As a first step, the MaxCombo suggests stronger evidence of a treatment difference (p=0.001) than LR test. As a next step, we looked into HR from Cox regression (estimate=0.78; 95\% CI (0.64, 0.95)) and KM medians (8.4 months (treatment) vs 6.9 months (control)) as key treatment effect summaries. As the PH assumption is doubtful from the diagnostics (Figure ~\ref{fig:EXSch}), we also looked into the RMST (estimate= 1.9; 95\% CI (0.39, 3.41)) and difference in month 12 milestone rates (estimate=0.09; 95\% CI (-0.02, 0.22)). Furthermore, the treatment effect is confirmed by the positive difference in milestone survival rates at early and late timepoints. Finally the piecewise HR plot shows a positive trend for survival benefit in favor of Pembrolizumab after first 3 months.

Finally for the PA3 trial, the LR test becomes statistically significant (one sided p-value =0.023) with a modest or clinically questionable treatment effect (HR= 0.834 with 95\% CI = (0.70, 0.99) or median difference of 0.33 months). The MaxCombo test fails to meet nominal statistical significance (p-value= 0.048) which is consistent with other tests (e.g., WKM, RMST). This is primarily due to the multiplicity adjustment for the different test statistic in MaxCombo.  However, caution should be used when applying the MaxCombo test if NPH is not anticipated.

The three examples above show the utility of the MaxCombo test and primary analysis approach when NPH is a possibility. It is evident that MaxCombo has benefit in case of delayed effect and crossing scenario survival scenarios. When the PH assumption is violated, the current regulatory standard of declaring a study positive or negative based on a single p-value (from the LR test) and estimating the treatment benefit using a single dual summary measure (HR from the Cox model) can be problematic. This suggests the value of a robust testing approach such as the MaxCombo and adaptive primary analysis strategy. It is also evident that  a single measure such as the HR is not adequate to describe the treatment benefit when PH assumption is violated. Additional measures such as piecewise HR, milestone survival and difference in RMST are essential for adequate description of treatment differences. 

\section{Design Approaches for Clinical Trials with Non-proportional Hazard}

In this section, we have discussed a design approach for a  confirmatory trial with the MaxCombo test. When NPH is a possibility, a confirmatory trial design needs to consider the uncertainty about the type of treatment effect. If a MaxCombo test is used for the primary testing, it is important that the sample size and total follow-up time of a trial ensure adequate power for the most likely treatment effect type under a reasonably conservative alternative hypothesis. Therefore, a carefully elicitation of the possible treatment effect type (e.g., delayed effect, crossing survival etc.) is important at the design stage. Often a confirmatory trial involves interim analysis for early stopping due to futility or overwhelming efficacy. Group sequential methods are popularly used in this context. A notable work regarding the use of group sequential design in trials with NPH includes ~\cite{LoganMo2015}. However, the key question is how to plan for interim analysis in a design with MaxCombo as primary analysis?  Although the group sequential strategies are well understood using LR test in the PH setting, little attention has been given to their performance when the effect of treatment varies over time. We present an interim stopping strategy when the MaxCombo test is the proposed primary testing.     

\subsection{Sample Size Calculation}

Under PH, the number of events determines the power of a design. But if the PH assumption is violated, enrollment rate, number of events, trial duration and total follow-up time play important roles in the power calculation. The final analysis timing based on the accumulation of events only may produce a design that finishes too early, is under-powered and failed to describe the impact of treatment over time. Therefore, a good design strategy with potential NPH needs to find a balance between the number of events (or sample size) and trial duration. A smaller sample size with fewer events but longer follow-up can provide more power and a better description of late behavior of the survival distribution than a larger sample size trial with short follow-up time. Unlike the LR test, a closed form expression for the sample size calculation and trial duration is not available for the MaxCombo test. Therefore, we propose a two-step approach for sample size calculation using an iterative procedure when the MaxCombo test is proposed for the primary testing of hypothesis. 

\begin{enumerate}
\item {\bf Determining minimum follow-up time (sample size time trade-off)}: First assume an enrollment duration and vary the minimum follow-up time for each patient after enrollment to optimize perceived trade-offs between sample size and trial duration. A general recommendation for minimum follow-up time is twice the median of control arm. However, this can vary for different settings. 
 	
\item {\bf Adjusted level of significance}: The MaxCombo test consists of four positively correlated Fleming-Harrington WLR test statistics. Therefore, the sample size calculation requires an adjusted significance level for each test to protect the overall type I error at a desired level (e.g. 2.5\%). We propose an adjusted level of significance calculation based on the asymptotic  multivariate distribution of $G^{0,0}$ , $G^{0,1}$, $G^{1,0}$, and $G^{1,1}$  (~\cite{Karrison2016}). This method requires the knowledge of correlation matrix between Fleming-Harrington WLR test statistics under the null hypothesis (equality of survival distribution). To estimate the correlation matrix, we simulate a trial with large sample size ($\geq$ 1000 per group) and with given enrollment and minimum follow-up time (determined in the previous step) for each patient. Here we generate data for treatment and control arms using piecewise exponential distributions with control event rate (under null). The correlation matrix is approximated by the empirical correlation matrix calculated from this large trial under the null hypothesis, given enrollment and follow-up time. This approach is efficient from a computational aspect and provides good estimate of the correlation matrix due to large sample theory. Finally, the statistical significance boundary for each component of the MaxCombo test is calculated by solving for a nominal Z-value of the multivariate distribution with mean zero and the correlation matrix resulting from the above simulation.  
	
\item {\bf Sample size calculation} After establishing the minimum follow-up time and adjusted level of significance, we use a minor modification of ~\cite{Hasegawa2014} to obtain the sample size and the required number of events for each component of the MaxCombo test at the adjusted significance level computed above. The minimum of these four numbers is chosen as the initial sample size and number of events. The final sample size, number of events and trial duration needs an iterative approach to confirm whether the design has type I error control and adequate power under alternative hypothesis scenarios of greatest interest (e.g., PH with minimal effect of interest and most conservative delayed effect alternative). 

If the operating characteristics are not adequate, the previous steps need to be repeated. Both the number of events and the minimum follow-up time need to be adjusted. This procedure continues until an acceptable power and type I error control are achieved.
\end{enumerate}     

\begin{figure}
	\includegraphics[width=\linewidth]{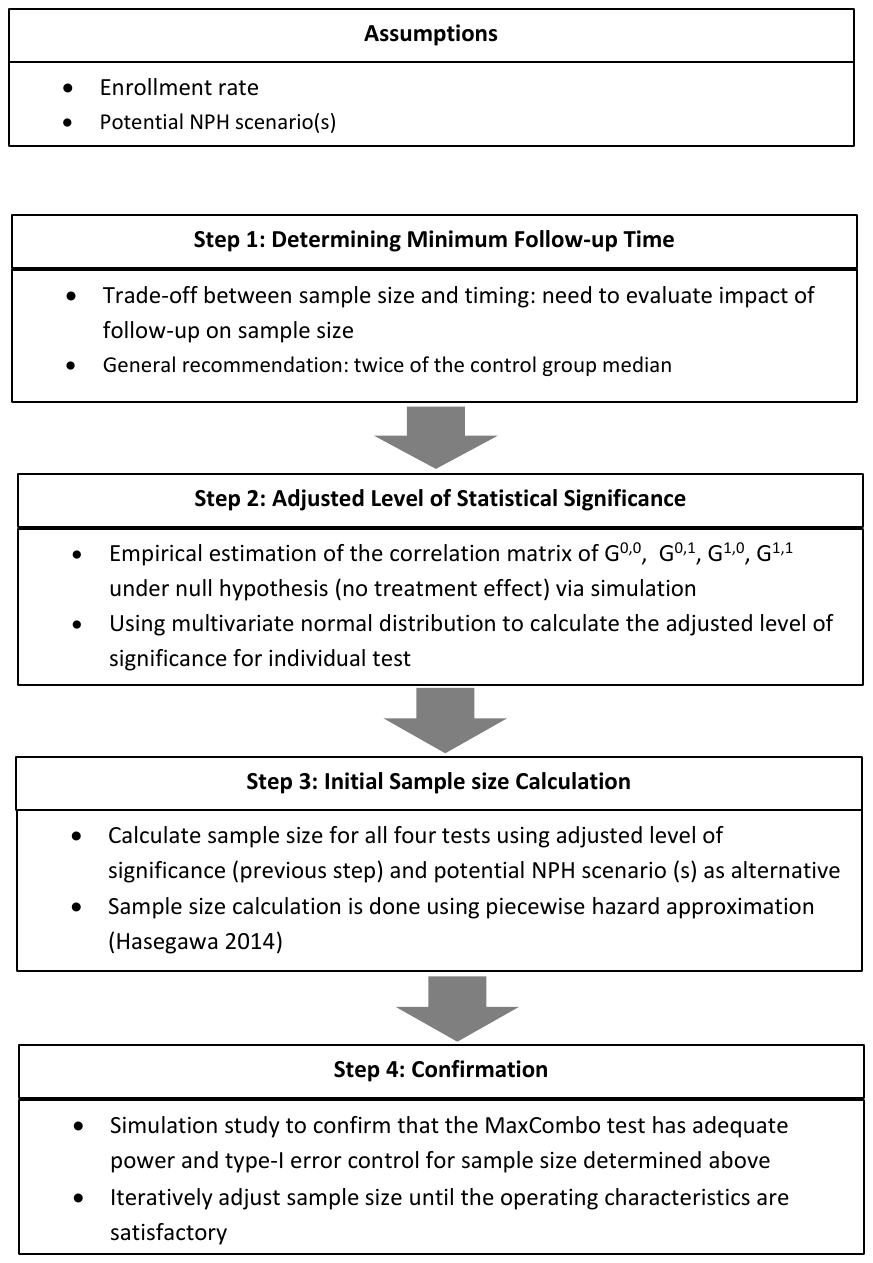}
	\caption{Overview of Sample Size Calculation}
	\label{fig:SS}
\end{figure}

The adjusted significance level is often twice that of an overly conservative Bonferroni correction, resulting in a smaller sample size requirement. In final analysis the p-value of MaxCombo test will be compared with the nominal p-value boundary (e.g., 2.5\% for one-sided p-value). For a confirmatory trial, we propose to include at least two treatment effect scenarios in the study protocol: a PH scenario and an expected NPH type for power calculation. Additional scenarios can also be considered based on the prior knowledge regarding the mechanism of action of the drug and possible NPH types. With the assumed enrollment rate,  minimum follow-up time, number of events and sample size calculated using the steps mentioned above, it is important to  demonstrate good operating characteristics in both the scenarios. This establishes the robustness of power for the MaxCombo test when the actual type of NPH is uncertain. More than one NPH scenario can be included based on the perceived needs at the design stage. A high level summary of the necessary steps for sample size calculation are provided in Figure ~\ref{fig:SS}.

\subsection{Interim Analysis}

Planning interim analysis requires a cautious approach when NPH is a possibility. Especially, for later emerging treatment effect scenarios (e.g., delayed effect or crossing survival) one needs to reconsider the traditional implementation of interim analyses. An early interim analysis will have smaller probability of stopping for efficacy and higher probability of crossing any futility bound under a delayed effect or crossing hazard scenario. On the other hand, if an interim analysis is too late, it may not be useful. While planning for interim analysis with potential NPH, it is important to find a balance between the risks of stopping too soon before late benefit emerges and the appropriately monitoring of the trial for futility. Thus, selection of interim analysis timing should consider both number of events and total follow-up. 

Available statistical literature regarding interim analysis with NPH is still small. A group sequential test for WLRT, WKM and combination of LR test \& Nelson Allen have been discussed by ~\cite{Hasegawa2016} and ~\cite{LoganMo2015} . In this paper,  we introduce a group sequential design strategy for planning interim analysis in a confirmatory trial when the primary testing is planned with the MaxCombo test. It incorporates both efficacy and futility bounds. For interim analysis, we propose using the LR test statistic. The three primary reasons for the recommendation are: i) to avoid the impact of shorter follow up time or trial duration in Fleming-Harrington WLR, ii) traditional interim boundaries based on the LR test are well accepted by regulatory authorities, and iii) using a more robust test at final analysis when data are mature to detect important benefits.  The final success boundary needs multiplicity adjustment due to the correlation between the LR test at interim and the MaxCombo test at the final analysis. We use the independent information increment assumption  (~\cite{Tsiatis1992}) and asymptotic multivariate distribution of interim LR test statistic and final MaxCombo test statistic to calculate the final analysis boundary. A detailed formula for the correlation matrix with one interim analysis and calculation of boundary for the final analysis are provided in appendix C. The final boundaries and correlation matrix are calculated based on actual number of observed event and follow-up. An example of the boundary calculation is provided in next subsection. 

Incorporating interim analysis for stopping in a design requires important statistical and operational considerations apart from the boundary calculation. This includes timing of interim analysis, probability of stopping, impact on the power, and overall benefit-risk etc. These aspects are out of scope for this paper. For timing of the efficacy interim, we recommend complete enrollment, accrual of at least 65-70\% of the planned events, and at least 6 months follow-up after last patient enrolled. An early futility analysis is problematic when treatment effect is emerging late. Therefore, we recommend statisticians not to perform futility analysis before 45-50\% of planned events are accrued unless there is a safety issue. The recommendation is to stop the trial if and only if the treatment seems harmful (e.g. HR $>$ 1.5).

Simulation plays an important role while designing a study with MaxCombo test. A structured approach for planning and reporting simulation studies is essential for successful execution in real-life trial.  Planning for simulation study includes data generating mechanism (underlying NPH scenarios expected), initial sample size (based on the components of MaxCombo), specification of the seed (to ensure reproducibility), number of iterations (to achieve appropriate precision). We suggest practitioners to consider both NPH and PH scenarios for data generation while planning for the protocol. The number of iteration needs to be large enough to provide appropriate precision. Half-width of the simulation 95\% confidence interval  is a well recognized measure of the precision for an estimate obtained via simulation. Based on our experience, a half-width of the simulation 95\% confidence interval up to 0.005 is generally acceptable in practice for type-I error assessment. However, the acceptable magnitude may be case specific.  Use of  computation approaches (e.g., state-of-art R packages like nphsim (~\cite{Keaven2018}) and simtrial (~\cite{simtrial})) , parallel computing systems) can increase the efficiency further. Finally, upfront discussion with regulators regarding simulation plan and preliminary results. Examples of design and analysis using nphsim and simtrial are provided in the supplementary materials. For further guidance, we refer to ~\cite{Morris2019} and ~\cite{Koehler2009}.

\subsection{Example}

In this subsection, we show an example of sample size calculation in a protocol for two arms (treatment vs control) randomized trial with a time to event endpoint. As mentioned above, when NPH is a possibility both trial duration and number of events are important components of the sample size section of the protocol. Furthermore, we assume that  based on the mechanism of treatment and evidence from early stage, there is a possibility for delayed effect. However, considerable uncertainty is associated with the occurrence of NPH and the lag time until treatment benefit emerges. This is a common phenomenon for immuno-oncology. 

We will assume 15 months of constant enrollment, a constant dropout rate of 0.001 per month, control group observations follow an exponential survival distribution with a median of 8 months. After careful elicitation of all available evidence, we assume a possible delayed treatment effect scenario (alternative hypothesis): no treatment effect for 6 months (HR=1), followed by a large treatment effect (HR= 0.56) thereafter. As this is a confirmatory trial for potential regulatory submission, strict control of 2.5\% type I error is required, and those investing in the study wish to ensure 90\% power under this NPH scenario. Below are the two steps for sample size calculation:  

\begin{enumerate}
	\item The first step is specification of the minimum follow up time for each patient or trial duration. We consider total trial duration of 18, 24, 32 and 40 months to compare required sample size for each component of the proposed MaxCombo test (Table ~\ref{tab:SSFollwps}). All sample sizes are calculated using the method of ~\cite{Hasegawa2014}.

\begin{table}[ht]
	\caption{Required Sample Size for Four Components of The MaxCombo Test under Different Trial Duration} \label{tab:SSFollwps}
	\centering
	\smallskip 
	\begin{threeparttable}
		\begin{tabular}{ccccc}
			                & \multicolumn{4}{c}{Number of Events/Sample Size}  \\
			 \cline{2-5} 
			                &             &             &            & \\      
			Trial duration  &$G^{0,0}$    &  $G^{0,1}$  &  $G^{1,0}$ & $G^{1,1}$\\ 
			   (months)     &             &             &            &\\  
			\hline
			18              & 1699/3160   &  709/1318   &  4028/7496 & 1002/1864\\
			24              & 755/1094    &  394/570    &  1829/2650 & 525/760\\
			32              & 511/628     &  296/364    &  1392/1712 & 399/490\\
			40              & 439/496     &  268/302    &  1329/1502 & 372/420\\
			\hline
		\end{tabular}
	\end{threeparttable}
\end{table}
	
	We note two things in the Table \ref{tab:SSFollwps}. First, the $G^{0,1}$ always results in the smallest sample size and event count requirement among the four WLR tests. Second, we select a study duration of 32 months (15 months enrollment + 17 months follow-up after last patient enrolled) given the steep increase in sample size for smaller trial duration and allowing more than twice the control arm median for minimum follow-up time.
	
	\item We generate the time to event data for 5000 patients (2500 per arm) using an exponential distribution with median 8 months in order to obtain a good estimate the correlation matrix of MaxCombo statistics under the null hypothesis. This sample size is large enough to generate a stable estimate. The correlation matrix for four Fleming-Harrington WLR test statistics is estimated using the empirical correlation from the large trial. Now using a grid search, the statistical significance boundary for MaxCombo is -2.286 with a nominal standard normal p-value of 0.011.  
	
	\item  The next step involves calculating the sample size and the number of events separately for the four components of MaxCombo test using the ~\cite{Hasegawa2014}. This calculation includes a) trial duration of 32 months, b) a drop-out rate of 0.001, c) specific delayed treatment effect alternative stated above, d) level of significance 1.1\% (calculated in step 2), and e) target power of 90\%. As the chosen NPH alternative reflects  delayed treatment effect, $G^{0,1}$ yields the minimum sample size and number of events. Therefore, the initial sample size of the trial is 442 with a target accrual of 360 events.   
	
	We confirm type I error and power using simulation under alternate hypotheses of interest. The initial sample size meets the type I error and power requirement based on 10000 simulations. We further assessed the power of MaxCombo under a PH scenario. The power of the proposed design under proportional hazards with HR=0.692 is 88.6\%. With 10000 simulations, the half width of the simulation 95\% confidence interval for type-I error (2.5\%) and power (90\% ) are 0.003 and 0.0058 respectively. One can consider running larger simulation (e.g., 40000) if more precision is desired.   
\end{enumerate}

Hence, the final analysis of the trial is planned after the accumulation of 372 events or 16 months after the last patient enrolled whatever happens last. The final sample size of the trial is 472. This represents considerable savings in sample size relative to the traditional LR test (690 patients and 544 events). Further details of the sample size calculation including correlation matrix for null distribution are provided in Appendix D.

Now, we add an interim analysis in the design to illustrate the calculation of analyses boundaries. We consider a single interim analysis for simplification. for example, an interim analysis is planned in this trial after a) enrollment is complete, b) at least 65\% events are observed, and c) at least 6 month of follow-up after enrollment is complete. The actual interim is performed after 270 events (75\% of the planned events). The interim boundary for efficacy is calculated by using the traditional alpha-spending function with O'Brien Fleming type boundary (~\cite{DemetsLan1994}). The interim efficacy boundary is -2.34 in Z-scale or  0.0096 in p-value scale. The final analysis boundary requires the correlation matrix between the interim LR test statistic and the four components of MaxCombo test under the null hypothesis. Lastly, we did a grid search of the final analysis cutoff that preserves the overall total type I error at 2.5\%.  The Z-value cutoff for final analysis is -2.305. Additional details are in Appendix D.

\section{Conclusion and Discussion}

The development plan for each experimental drug is unique. Current design and analysis approaches of a confirmatory clinical trial with a time to event endpoint depend heavily on the proportional hazards assumption which is questionable in many situations. Traditional approaches are less efficient when the treatment effect is not constant. Non-constant effects like delayed effect or crossing survival, are often observed. It is important to propose a flexible and robust primary analysis strategy to cope with possible treatment effect patterns. Many authors have proposed statistical methods for analyzing clinical trials when the PH assumption is violated. However, none of them are robust enough to have adequate power for wide number of NPH scenarios (~\cite{NPHsim2018}) which is critical in practice due to the uncertainty related to possible NPH types at design stage. 

We propose a robust statistical approach for primary hypothesis testing in a confirmatory clinical study with a time to event endpoint. The proposed methodology is flexible and adaptive enough to provide good statistical properties under PH and different types of NPH. The proposed MaxCombo test is a combination of four Fleming-Harrington WLR tests which can handle different treatment effect patterns. The operating characteristics (~\cite{NPHsim2018}) including type-I error control and illustrative examples demonstrate the utility of the MaxCombo test. The test shows clear advantages when a large treatment benefit emerges later on in the trial (PH with marginal effect, delayed effect and crossing hazard scenario in Section 2.3). The PH scenario (digitized PA3 trial data) shows a possible downside of the MaxCombo test as a marginally positive LR test in a case that reflects proportional hazards would have been a close miss if the MaxCombo test had been utilized. While one could argue that the survival benefit was minimal, it was an overall survival benefit for pancreatic cancer, an indication where the authors point out that progress in treatment options has been slow. The ultimate value of the treatment for patients may depend on trade-offs between toxicity and secondary efficacy endpoints. For trial designers, this should point out that the MaxCombo test might not be optimal if there is a strong reason to believe that the benefit from a new treatment will be immediate and sustained. This assumption generally would not be made for immunotherapies or, say, chronic treatment of diabetes or lipids to reduce long-term cardiovascular risk. Recently MaxCombo test has been used as a sensitivity analysis in a randomised, open-label, multicentre, phase 3 trial in previously untreated patients with unresectable, locally advanced or metastatic urothelial carcinoma (~\cite{POWLES2020}).

Following the recent criticism about the WLR tests (~\cite{MagirrBurman2018}, ~\cite{FreidlinKorn2019}), we have performed additional simulations to evaluate the MaxCombo test under some extreme scenarios such as strong null  (~\cite{MagirrBurman2018}, ~\cite{FreidlinKorn2019}) and one severe crossing hazard. The simulation shows a high probability of rejecting null hypothesis for the MaxCombo test under strong null 2 (~\cite{FreidlinKorn2019}). However, such scenarios can be handled with the modified MaxCombo test.  Note that, these scenarios (specially strong null 2 which has an early hazard ratio of 16 that would result in a very large early excess mortality) are unlikely in real life and  will result in the trial getting stopped early by a data monitoring committee (DMC) due to the safety concerns.
 
We have proposed a stepwise approach for reporting the treatment effect quantifiers depending on the validity of PH. Though HR may be sufficient under PH, a single summary measure is not adequate when the treatment benefit is not constant over time. We recommend trialists report additional measures like milestone survival rates and RMST as primary treatment effect quantifiers when the PH assumption is questionable. This is different from current practice. The adaptive nature of the proposed approach helps to choose the most appropriate summary measures for describing the treatment effect. This provides statisticians the required flexibility while being compliant with ICH E9 guidance (~\cite{ICHE9}). When PH is violated and a difference is established using the MaxCombo test, additional measures such as milestone survival, difference in RMST, WHR and piecewise HR are useful in interpreting the trial results. There is a possibility that the WHR overestimates treatment effect in some situations by allowing high weights on the tail events. ~\cite{Bartlett2020} also showed concerns about using HR or WHR due to the lack of causal interpretation. Therefore, the other supportive measures are important to understand the complete picture (~\cite{LinL2020}). However, a traditional summary measure such as HR may be adequate  when the PH assumption is reasonable.

The second part of the paper introduces a design approach for confirmatory trials with the MaxCombo test. We have developed a stepwise and iterative approach for calculating sample size when the final analysis is based on MaxCombo test. In contrast to the LR test, the design approach with the MaxCombo test needs to consider both the number of events and total follow up time for good statistical operating characteristics (type I error and power). The study protocol needs to state all the requirements along with a detailed simulation plan for assessing the design operating characteristics. We have provided a detailed example and R codes to help practitioners. The design approach uses the asymptotic joint distribution of four Fleming-Harrington WLR tests which is more efficient than other conservative multiplicity adjustments or LR test alone. We have also proposed a simple yet effective interim analysis procedure for early stopping for efficacy. As adequate closed form formulas are not available, simulation plays an important role at the design stage.  Efficient R packages (e.g. nphsim (~\cite{Keaven2018}) or simtrial (~\cite{simtrial}))  can be useful in this context. 

Finally, in this paper we have discussed possible analysis and design for a confirmatory trial.  The proposed approach is intended to improve study design and appropriate discussion about advancement of new therapies that may currently not be considered due to overly restrictive expectations on statistical testing.

\begin{acks}
	We would like to thank all the members of the cross-pharma non-proportional hazards working group for their input. Special thanks to Renee B Iacona (Astrazeneca Pharmaceuticals), Tai-Tsang Chen (Bristol-Myers Squibb Company), Ray Lin (Roche), Ji Lin (Eli Lilly \& Co.), Tianle Hu (Eli Lilly \& Co.) for their contributions and constant support. We would also like to thank Dr. Susan Halabi from Duke University for her valuable suggestion and encouragements. A special thanks goes to the referees and Associate editor (AE) and editors of Statistics in Biopharmaceutical Research. The comments from referee and AE helped significantly in improving the contents of this paper. Finally, we would also thank the industry and the FDA participants in the Duke-Margolis workshop for their valuable input and discussions. All materials of the Duke-Margolis workshop are available at this  \href{http:/healthpolicy.duke.edu/events/public-workshop-oncology-clinical-trials-presence-non-proportional-hazards}{link}.
\end{acks}

\bibliographystyle{ECA_jasa}
\bibliography{NPHreference}

\newpage

\section*{Appendix A: Combination test and Calculation of p-value}
The proposed combination test 
\begin{eqnarray*}
Z_{max} &=& max\{G^{\rho_{i},\gamma_{j}}: \; (\rho_{i},\gamma_{j}) = (0,0), (0,1), (1,0),(1,1)\}
\end{eqnarray*}

Using the result from ~\cite{Karrison2016}, the asymptotic null distribution of four WLR test statistics follows a multivariate distribution  with mean {\boldmath $0$} and correlation matrix {\boldmath $\Gamma$}: 
\begin{eqnarray*}
(G^{0,0}, G^{0,1}, G^{1,0}, G^{1,1}) \sim N_{4}(\mbox{\boldmath{$0$}},\; \mbox{\boldmath{$\Gamma$}})
\end{eqnarray*}

With correlation matrix {\boldmath $\Gamma$}=$((\eta_{ij}))$ is of the following form;
\begin{eqnarray*}
\eta_{ij} &=& \frac{\hbox{Cov}(G^{\rho_{i},\gamma_{i}}, G^{\rho_{j},\gamma_{j}})}{\sqrt{V(G^{\rho_{i},\gamma_{i}})V(G^{\rho_{j},\gamma_{j}})}} = \frac{V(G^{\frac{\rho_{i} + \rho_{j}}{2},\frac{\gamma_{i}+\gamma_{j}}{2}})}{\sqrt{V(G^{\rho_{i},\gamma_{i}})V(G^{\rho_{j},\gamma_{j}})}} \;\;\;\; \mbox{for}\; i \neq j \\
&=& 1 \;\;\;\; \mbox{for}\; i = j
\end{eqnarray*}

Therefore, the one-sided p-value of MaxCombo test is calculated using a multivariate normal calculation given below:
\begin{eqnarray*}
P(Z_{max} > z_{max}|H_{0}) &=& P(\max(G^{0,0}, G^{0,1}, G^{1,0}, G^{1,1}) > z_{max}| H_{0}) \\	
                           &=& 1-\int_{-\infty}^{z_{max}}\int_{-\infty}^{z_{max}}\int_{-\infty}^{z_{max}}\int_{-\infty}^{z_{max}}\phi_{4}(\mbox{\boldmath{$\omega$}}, \mbox{\boldmath{$0$}}, \mbox{\boldmath{$\Gamma$}})d\mbox{\boldmath{$\omega$}}
\end{eqnarray*}	
	
$z_{max}$ is the observed value of MaxCombo test statistic and $\phi$ is the pdf of 4-dimensional multivariate distribution. The power calculation of MaxCombo does not have a closed expression. Therefore, a simulation approach is required for power calculation. Supplementary material contains an example R codes for calculating p-value of the MaxCombo test and corresponding adjusted 95\% confidence interval using nphsim package.

\section*{Appendix B: Calculation of Simultaneous Confidence Interval}
Let, $HR^{MaxCombo}$ is the estimated WHR using the best weight as per MaxCombo and using weighted Cox regression (~\cite{Therneau2000}). Therefore, a $100\times(1-\alpha)\%$ simultaneous confidence interval corresponding for WHR related to MaxCombo can be calculated as $HR^{MaxCombo} \; \pm \; C^{*}\times SE(HR^{MaxCombo})$. $SE(HR^{MaxCombo})$ is the standard error of $HR^{MaxCombo}$ and $C^{*}$ is calculated using the asymptotic multivariate normal distribution of WHR (~\cite{Karrison2016}). 

\section*{Appendix C: Calculating interim and final boundaries with one interim analysis}

Let $t$ be the fraction of events at interim over planned event count at final analysis. $G^{0,0}(t)$, $G^{0,0}(1)$, $G^{0,1}(1)$, $G^{1,0}(1)$, and $G^{1,1}(1)$ are the LR test statistic at the interim analysis and test statistics of the MaxCombo test at the final analysis (LR test, (FH(0,0), FH(0,1), FH(1,0), and FH(1,1) respectively). 
\begin{eqnarray*}
Z_{max} = max\{G^{0,0}(1), G^{0,1}(1), G^{1,0}(1), G^{1,1}(1)\}
\end{eqnarray*}

Now, the correlation between $G^{0,0}(t)$ and $G^{\rho,\gamma}(1)$; $\rho, \gamma =0,1$ can be calculated using independent increment (\cite{Tsiatis1992}) and  ~\cite{Karrison2016} 
\begin{eqnarray*}
cov(G^{0,0}(t), G^{\rho,\gamma}(1)) &= & cov(G^{0,0}(t), G^{\rho,\gamma}(t))\;\; \mbox{assuming independent increment}\\
&=& var(G^{\frac{\rho}{2},\frac{\gamma}{2}}(t))   
\end{eqnarray*}

Therefore, the correlation between $G^{0,0}(t)$ and $G^{\rho,\gamma}(1)$ 
\begin{eqnarray*}
corr(G^{0,0}(t), G^{\rho,\gamma}(1)) = \frac{V(G^{\frac{\rho}{2},\frac{\gamma}{2}}(t))}{\sqrt{V(G^{0,0}(t))\times V(G^{\rho,\gamma}(1))}}
\end{eqnarray*}

The boundary for the final analysis ($z^{F}_{max}$) must satisfy the equation below;
\begin{eqnarray*}
P(Z_{I} > z_{I}| H_{0}) + P(Z_{I} \leq z_{I}, Z^{F}_{max} > z^{F}_{max} | H_{0}) \leq 0.025 
\end{eqnarray*} 

Here $Z_{I}$, $z_{I}$, and $Z^{F}_{max}$ are LR test statistic at interim analysis, interim efficacy boundary, and MaxCombo test statistic for final analysis. $z_{I}$ is determined by well-known spending functions (e.g., O'Brien-Fleming) and $z^{F}_{max}$ can be solved using a grid search. 

\section*{Appendix D: Details of Example in Section 3.3}

\begin{enumerate}

\item[I]  \underline{Details of Sample Size Calculation without Interim Analysis}

The estimated correlation matrix of the null distribution based on simulation steps described in Section 3.3

\begin{eqnarray*}
\mbox{\boldmath{$\Gamma_{H_{0}}$}}	= \begin{bmatrix}
	1.000  & 0.864  & 0.913 & 0.940  \\ 
	0.864  & 1.000  & 0.583 & 0.892 \\ 
	0.913  & 0.583  & 1.000 & 0.792 \\ 
	0.940  & 0.892 & 0.793 & 1.000\\  
	\end{bmatrix}
\end{eqnarray*}

Therefore, the boundary for MaxCombo ($Z_{cutoff}$) is obtained by solving the equation below;
\begin{eqnarray*}
	\Phi_{4}(Z_{cutoff}, \mbox{{\boldmath $0$}}, \mbox{\boldmath{$\Sigma_{H_{0}}$}}) \leq 0.025
\end{eqnarray*}
$\Phi_{4}$ is the CDF of 4-dimensional multivariate normal distribution. This yields $Z_{cutoff}$ = -2.286. Now, the calculation of sample size for four components of the MaxCombo test uses a) level of significance 1.11\%, b) target power of 90\%, c) constant enrollment for 15 months, d) follow-up for 17 months after last patient enrolled, d) dropout rate 0.001 patients per month, and d) alternative hypothesis: no treatment effect (HR=1) for first 6 month followed by clinically meaningful treatment effect (HR=0.56). The sample size for each test is given in Table ~\ref{tab:NPH:minSS} below:

\begin{table}[ht]
	\caption{Sample Size for Four Components of the MaxCombo Test} \label{tab:NPH:minSS}
	\centering
	\smallskip 
	\begin{tabular}{ccc}
		\hline
		Test      & Sample Size & Event \\
		\hline
		$G^{0,0}$ &   828       &  653 \\
		$G^{0,1}$ &   472       &  372 \\
		$G^{1,0}$ &  2190       & 1726 \\
		$G^{1,1}$ &   630       &  497\\
		\hline
	\end{tabular}
\end{table}

Therefore,  the final analysis of the trial is planned after the accumulation of 372 events or 16 months after the last patient enrolled whatever happens last. The final sample size of the trial is 472.   

\item[II] \underline{Details of Interim and Final Analysis Boundary Calculation}

The correlation matrix for interim LR test statistic and four Fleming Harrington WLR test statistic for final analysis is 
\begin{eqnarray*}
\mbox{\boldmath{$\Gamma_{H_{I}}$}}	= \begin{bmatrix}
	1.000 & 0.858 & 0.565 & 0.926 & 0.769 \\
	0.858 & 1.000 & 0.863 & 0.930 & 0.940 \\
	0.565 & 0.863 & 1.000 & 0.617 & 0.922 \\
	0.926 & 0.930 & 0.618 & 1.000 & 0.794 \\
    0.768 & 0.940 & 0.922 & 0.794 & 1.000\\  
\end{bmatrix}
\end{eqnarray*}
Therefore the final boundary ($z_{F}$) can be calculated using a grid search to solve for 

\end{enumerate}

\begin{eqnarray*}
	P(Z_{I} < -2.34| H_{0}) + P(Z_{I} > -2.34, M_{F} < z_{F} | H_{0}) \leq 0.025 
\end{eqnarray*} 
The interim stopping boundary ($z_{I}$ = -2.34) is calculated using the Lan-DeMets (O’Brien-Fleming) $\alpha$-spending function. Supplementary material contains an example R code for sample size and interim analysis boundary calculation.

\end{document}